\documentclass[11pt]{article}
\newcommand{\Atr}{A(t,r)}
\newcommand{\Btr}{B(t,r)}
\newcommand{\Ar}{A(r)}
\newcommand{\Br}{B(r)}
\newcommand{\Atrsq}{A^{2}(t,r)}
\newcommand{\Btrsq}{B^{2}(t,r)}

\newcommand{\betatx}{\beta(\tau,\chi)}
\newcommand{\alphatxsq}{\alpha^{2}(\tau,\chi)}
\newcommand{\betatxsq}{\beta^{2}(\tau,\chi)}
\newcommand{\alphat}{\alpha(\tau)}
\newcommand{\betat}{\beta(\tau)}
\newcommand{\betatdot}{\dot{\beta}(\tau)}
\newcommand{\alphatdot}{\dot{\alpha}(\tau)}

\newcommand{\betatdotdot}{\ddot{\beta}(\tau)}
\newcommand{\alphatinv}{\alpha^{-1}(\tau)}
\newcommand{\betatinv}{\beta^{-1}(\tau)}
\newcommand{\alphatsq}{\alpha^{2}(\tau)}
\newcommand{\betatsq}{\beta^{2}(\tau)}
\newcommand{\alphatcubed}{\alpha^{3}(\tau)}
\newcommand{\st}{\sin\theta}
\newcommand{\ct}{\cos\theta}
\newcommand{\stsq}{\sin^{2}\theta}

\newcommand{\sph}{\sin\phi}
\newcommand{\cph}{\cos\phi}
\newcommand{\tpg}{t_{\mbox{\tiny{pg}}}}
\newcommand{\stuff}{\left\{\left[\Ar\Br\right]^{-2} - P_{0}B^{-2}(r)\right\}^{1/2}\, B^{2}(r)}
\newcommand{\alphao}{\alpha_{(0)}}

\newcommand{\dotalphao}{\dot{\alpha}_{(0)}}
\newcommand{\dotdotalphao}{\ddot{\alpha}_{(0)}}
\newcommand{\dotdotdotalphao}{\dddot{\alpha}_{(0)}}
\newcommand{\dotdotdotdotalphao}{\ddddot{\alpha}_{(0)}}
\newcommand{\betao}{\beta_{(0)}}
\newcommand{\dotbetao}{\dot{\beta}_{(0)}}
\newcommand{\dotdotbetao}{\ddot{\beta}_{(0)}}
\newcommand{\dotdotdotbetao}{\dddot{\beta}_{(0)}}
\newcommand{\dotdotdotdotbetao}{\ddddot{\beta}_{(0)}}

\newcommand{\ppar}{p_{\mbox{\tiny{$\parallel$}}}}
\newcommand{\pperp}{p_{\mbox{\tiny{$\perp$}}}}
\newcommand{\Df}{\frac{df(T)}{dT}}
\newcommand{\Ddf}{\frac{d^{2}f(T)}{dT^{2}}}
\usepackage{times,fancyhdr}
\usepackage{latexsym}
\usepackage{amssymb}
\usepackage{amsmath}
\usepackage{graphicx}
\usepackage{bigfoot}
\usepackage{bm}
\usepackage{sectsty}
\usepackage{dsfont}

\newcommand*{\Scale}[2][4]{\scalebox{#1}{$#2$}}%

\sectionfont{\normalsize}
\subsectionfont{\small}
\subsubsectionfont{\mdseries\small\underline}

\date{}

\title{{\normalsize{\bf{MATTER CONDITIONS FOR REGULAR BLACK HOLES IN $\mathbf{f(T)}$ GRAVITY}}}} 
\author{{\small Joshua Aftergood \footnote{jaftergo@sfu.ca}} \\
\it{\small Department of Physics, Simon Fraser University,} \\
\it{\small Burnaby, British Columbia, Canada, V5A 1S6} \\[-0.1cm]
\line(1,0){45}\\
{\small Andrew DeBenedictis \footnote{adebened@sfu.ca}} \\
\it{\small Department of Physics} \\
\it{{\footnotesize{and}}}\\
\it{\small The Pacific Institute for the Mathematical Sciences,}\\
\it{\small Simon Fraser University, Burnaby, British Columbia, Canada, V5A 1S6}\\[-0.1cm]
}

\begin{document} 

\pagestyle{fancy}
\fancyhead{} 
\fancyhead[OR]{\thepage}
\fancyhead[OC]{{\small{REGULAR BLACK HOLES IN ${f(T)}$ GRAVITY}}}
\fancyfoot{} 
\renewcommand\headrulewidth{0.5pt}
\addtolength{\headheight}{2pt} 

\maketitle 
\vspace{-1.0cm}
\begin{center}
\rule{175pt}{0.5pt}
\end{center}

\begin{center}
 {\footnotesize{\bf{ABSTRACT}}}
\end{center}
\noindent We study the conditions imposed on matter to produce a regular (non-singular) interior of a class of spherically symmetric black holes in the $f(T)$ extension of teleparallel gravity. The class of black holes studied (T-spheres) is necessarily singular in general relativity. We derive a tetrad which is compatible with the black hole interior and utilize this tetrad in the gravitational equations of motion to study the black hole interior. It is shown that in the case where the gravitational Lagrangian is expandable in a power series $f(T)=T+\underset{n\neq 1}{\sum} b_{n}T^{n}$ that black holes can be non-singular while respecting certain energy conditions in the matter fields. Thus the black hole singularity may be removed and the gravitational equations of motion can remain valid throughout the manifold. This is true as long as $n$ is positive, but is not true in the negative sector of the theory. Hence, gravitational $f(T)$ Lagrangians which are Taylor expandable in powers of $T$ may yield regular black holes of this type. Although it is found that these black holes can be rendered non-singular in $f(T)$ theory, we conjecture that a mild singularity theorem holds in that the dominant energy condition is violated in an arbitrarily small neighborhood of the general relativity singular point if the corresponding $f(T)$ black hole is regular. The analytic techniques here can also be applied to gravitational Lagrangians which are not Laurent or Taylor expandable.
\par
\vspace{0.2cm}\noindent{\small{\textbf{PACS numbers:} 04.20.Dw,\;04.70.-s,\; 04.50.Kd}}
\vskip -0.1cm
\vspace{\baselineskip}\hrule

\section{{Introduction}}\label{S:intro}
There has been extensive study in the literature of the peculiar and not so peculiar properties of black holes over the years since the advent of general relativity. The Schwarzschild metric \cite{ref:schw}, one of the earliest non-trivial solutions to Einstein's field equations, possessed at least two peculiarities. One occurs at what is now known as the event horizon and was later discovered to be rather benign. The other occurs at what seemed to be the ``center'' of the solution and was later seen as more serious as it seemed to predict the breakdown of all physics at the location of this singularity. Since the early studies there have been many attempts at finding ways to alleviate this singularity with, for example, the addition of matter, as any astrophysical black hole would form from gravitational collapse. In 1965 and 1970 Penrose and Hawking introduced their now famous singularity theorems \cite{ref:pen} \cite{ref:penhawk}, which in the context of black holes essentially state that if a horizon forms, and the matter obeys certain reasonable conditions (such as the strong energy condition) then general relativity predicts that a singularity will be present somewhere in the spacetime. Other studies concerned with singularity resolution involve candidate theories of quantum gravity \cite{ref:bojosing}, \cite{ref:adebsing}, the possibility of various matter fields with exotic properties (which perhaps only manifest in the high curvature regime) (see \cite{ref:nonlinembh}-\cite{ref:klinkhamer} and references therein), and alternative theories of gravity which possess an observationally compatible weak field limit, but deviate from Newtonian gravity and General Relativity under more extreme conditions (\cite{ref:dilaton}-\cite{ref:palatini} and references therein). It is in the vein of this last possibility that we present the study here and find general conditions which yield a regular black hole in teleparallel gravity in a situation where none would be found in General Relativity. 

One alternative to the curvature based general relativity and its various extensions is a theory of gravity which is purely torsion based. These theories have experienced a bit of a renaissance in the past decade. Specifically, the extended teleparallel theory is one which has garnered much attention since it produces a theory equivalent to general relativity in an appropriate limit ($f(T)=T$, \cite{ref:tpoverview}) and also retains second-order equations of motion \cite{ref:telebook}. In this theory the fundamental object is the tetrad, $h^{a}_{\;\mu}$, and the action is given by
\begin{equation}
S=\frac{1}{16\pi}\int \left\{f(T)+\mathcal{L}_{\mbox{\tiny{m}}}\right\}\,\mbox{det}[h^{a}_{\;\mu}]\,d^{4}x\,. \label{eq:gravact}
\end{equation}
Here $\mathcal{L}_{\mbox{\tiny{m}}}$ represents the matter Lagrangian density. The indices employed in this paper have the following convention: Unadorned Greek indices are spacetime indices, whereas Latin indices span the local tangent spacetime of the gravitational degrees of freedom. In what follows we also occasionally project quantities into a local orthonormal frame which is \emph{not} related to the tetrad of the gravitational degrees of freedom\footnote{Although one can choose the local orthonormal frame to coincide with the tetrad of the gravitational field's degrees of freedom, it is generally not convenient to do so.}. The indices representing the local orthonormal frame are hatted Greek indices. The quantity $T$ is the torsion scalar which is defined via a linear combination of quadratic contractions of the torsion tensor, $T_{\rho\mu\nu}$:
\begin{equation}
T=\frac{1}{4}T^{\rho\mu\nu}T_{\rho\mu\nu}+ \frac{1}{2}T^{\rho\mu\nu}T_{\nu\mu\rho}-T_{\rho\mu}^{\;\;\;\rho}T^{\nu\mu}_{\;\;\;\;\nu}\,. \label{eq:T}
\end{equation}
The torsion tensor itself is defined via the commutator of the curvature-less Weitzenb\"{o}ck connection, $\Gamma^{\lambda}_{\;\;\mu\nu}$: 
\begin{equation}
T^{\lambda}_{\;\;\mu\nu}=\Gamma^{\lambda}_{\;\;\nu\mu}-\Gamma^{\lambda}_{\;\;\mu\nu}= h^{\lambda}_{\;a}\left(\partial_{\mu}h^{a}_{\;\nu}-\partial_{\nu}h^{a}_{\;\mu}\right)\,. \label{eq:torsion}
\end{equation}
The torsion scalar (\ref{eq:T}) differs from the Ricci scalar by a total divergence, and hence General Relativity is recovered in the limit that $f(T) \rightarrow T$. For this reason we will always include a term linear in $T$ in the subsequent analysis and specifically will consider functions of the form
\begin{equation}
f(T)=T+\underset{n\neq 1}{\sum} b_{n}T^{n}\,, \label{eq:ourf}
\end{equation}
with the $b_{n}$ constants. 

Variation of the action (\ref{eq:gravact}) with respect to the tetrad yields the gravitational equations of motion,
\begin{equation}
\Scale[0.95]{\left[h^{-1} h^{a}_{\;\mu} \partial_{\rho}\left(h\, h_{a}^{\;\lambda} S_{\lambda}^{\;\nu\rho}\right) + T^{\alpha}_{\;\lambda\mu} S_{\alpha}^{\;\nu\lambda}\right]\frac{df(T)}{dT} + S_{\mu}^{\;\nu\lambda}\,\partial_{\lambda}T \left(\frac{d^{2}f(T)}{dT^{2}}\right) +\frac{1}{4} \delta^{\nu}_{\;\mu}\,f(T)= 4\pi\mathcal{T}^{\;\nu}_{\mu}}\,, \label{eq:eoms}
\end{equation}
where $h$ is the determinant of the tetrad and $S_{\lambda}^{\;\nu\rho}$ is the modified torsion tensor:
\begin{equation}
 S_{\lambda}^{\;\nu\rho}=\frac{1}{2}\left(K^{\nu\rho}_{\;\;\;\lambda} + \delta^{\nu}_{\;\lambda} T^{\sigma\rho}_{\;\;\;\sigma}-\delta^{\rho}_{\;\lambda} T^{\sigma\nu}_{\;\;\;\sigma}\right)\,, \label{eq:modtor}
\end{equation}
and $K^{\nu\rho}_{\;\;\;\lambda}$ the contorsion tensor:
\begin{equation}
 K^{\nu\rho}_{\;\;\;\lambda}=\frac{1}{2}\left(T^{\rho\nu}_{\;\;\;\lambda}+T_{\lambda}^{\;\nu\rho}-T^{\nu\rho}_{\;\;\;\lambda}\right)\,. \label{eq:contorsion}
\end{equation}
$\mathcal{T}^{\;\nu}_{\mu}$ is the usual stress-energy tensor.

One issue present in teleparallel gravity which is not present in curvature theories where the metric is the fundamental gravitational object, such as $f(R)$ theories, is that the action (\ref{eq:gravact}) is not \emph{locally} Lorentz invariant. That is, two different tetrads, related to each other via a local Lorentz transformation, will yield physically distinct equations of motion. Therefore, a certain metric (which still defines the causal structure of the theory) is compatible with many tetrads all of which yield inequivalent relations between the matter content and the gravitational field via the field equations. Out of this freedom one must find a tetrad which yields ``good'' equations of motion. Reasonable criteria for a good tetrad are:
\begin{itemize}
\item The tetrad chosen should not restrict the form of $f(T)$ \cite{ref:goodbadtets}. That is, the tetrad needs to retain acceptable equations of motion regardless of the function $f(T)$ and not just work well for certain functions. Our minimal conditions for acceptable equations of motion are the following.
\item The tetrad must produce equations of motion which are compatible with a symmetric stress-energy tensor $\mathcal{T}_{\mu\nu}=\mathcal{T}_{\nu\mu}$.
\item The resulting equations of motion should not produce peculiar physics. For example, in spherical symmetry there should be no energy transport in the angular directions. In a homogeneous scenario, there should be no energy flux from one location to the other, etc.
\end{itemize}
It turns out it is generally non-trivial to find a tetrad that satisfies these conditions. In the realm of cosmology, for example, an array of frames have been analyzed in \cite{ref:nontrivframes}. We discuss this situation for the case of spherically symmetric black hole interiors in section \ref{S:tetrad}. In section \ref{S:singularity} we try to identify what presents a physical singularity in teleparallel theory and derive the criteria required to eliminate this singularity. We also show that it is possible to eliminate the singularity with matter which obeys energy conditions at the general relativity singular point, and conjecture that the dominant energy condition is violated in an arbitrarily small neighborhood about the singular point (although it does not need to be violated exactly at this point). We conclude that regularizing the black hole  is possible for the addition of terms with any positive value of $n\geq2$ in (\ref{eq:ourf}) tested, but for negative $n$ regularization fails and the situation is the same as in general relativity. Finally, in section \ref{S:conclusion} we summarize the results and provide some concluding statements.

\section{{A suitable tetrad}}\label{S:tetrad}
The line element appropriate for a spherically symmetric, not necessarily vacuum, black hole exterior can be cast in the form
\begin{equation}
 ds^{2}=\Atrsq\,dt^{2}-\Btrsq\,dr^{2}-r^{2}\,d\theta^{2}-r^{2}\stsq\,d\phi^{2}\,, \label{eq:rdomline}
\end{equation}
with a horizon present where $\Atr=0$. The obvious tetrad to use for such a line-element is the diagonal one:
\begin{equation}
\left[h^{a}_{\;\mu}\right]_{\mbox{\tiny{diag}}}=\left( \begin{array}{cccc}
\Atr & 0 & 0 & 0\\
0 & \Btr & 0 & 0 \\
0 & 0 & r & 0 \\
0 & 0 & 0 & r\sin\theta  \end{array} \right)\,. \label{eq:rdomdiagtet}
\end{equation}
However, it is well-known that this tetrad does not produce acceptable equations of motion when utilized in (\ref{eq:eoms}) \cite{ref:tamaninipres}, unless $f(T)=T$. For example, for $f(T)\neq T$ this tetrad can produce off-diagonal components to the stress-energy tensor even when the $t$ dependence is not present and there should be no momentum flux.

Note however that one may perform Lorentz transformations on the Lorentz index of the tetrad, which as briefly discussed in the introduction, will alter the equations of motion if the transformation is a local one. In the literature a common tetrad that is used, which produces acceptable equations of motion is the following rotated tetrad:
\begin{equation}
\left[h^{a}_{\;\mu}\right]_{\mbox{\tiny{rot}}}=\left( \begin{array}{cccc}
\Ar & 0 & 0 & 0\\
0 & \Br\st\cph & r\ct\cph & -r\st\sph \\
0 & \Br\st\sph & r\ct\sph & r\st\cph \\
0 & \Br\ct & -r\st & 0  \end{array} \right)\,. \label{eq:rdomrottet}
\end{equation}
This tetrad has been utilized in studies of charged black hole exterior spacetimes \cite{ref:chargedbh} and spherically symmetric stars \cite{ref:ftstars}, \cite{ref:anisonondiag}, and a more generalized version used in \cite{ref:specialsphsymsol}. A novel tetrad was utilized in \cite{ref:statsphsols} to study a class of static, spherically symmetric solutions. The diagonal tetrad (\ref{eq:rdomdiagtet}) was utilized in \cite{ref:ddimchargedbh} to study higher dimensional models. (\ref{eq:rdomrottet}) has also been used to discern some torsion only effects with non-minimal coupling to scalar fields in the teleparallel equivalent of general relativity \cite{ref:dubravko}. Dirac field coupling has been considered in \cite{ref:nonmindiraccoupling}. A tetrad yielding a vacuum Schwarzschild solution for an array of Lagrangians has been introduced in \cite{ref:spheresymspacetimes}. Also, the Kerr solution has been studied in the teleparallel equivalent of general relativity utilizing a tetrad appropriate for that spacetime in \cite{ref:reappraisal}.

The tetrad (\ref{eq:rdomrottet}) works well for systems where the line element is given by (\ref{eq:rdomline}) but for black hole interiors this tetrad is not suitable for $f(T)\neq T$ due to a somewhat subtle reason. A black hole interior's line element can be cast as:
\begin{equation}
 ds^{2}=\alphatxsq\,d\tau^{2}-\betatxsq\,d\chi^{2}-\tau^{2}\,d\theta^{2}-\tau^{2}\stsq\,d\phi^{2}\,, \label{eq:tdomline}
\end{equation}
and a horizon exists where $\betatx=0$. In this coordinate chart (sometimes called the T-domain chart, T indicating the time dependence of the interior spacetime, \emph{not} the torsion scalar) the Schwarzschild black hole interior would read
\begin{equation}
 ds^{2}_{\mbox{\tiny{Schw}}}= \frac{d\tau^{2}}{\frac{2M}{\tau}-1} - \left(\frac{2M}{\tau}-1\right)\,d\chi^{2} -\tau^{2}\,d\theta^{2} -\tau^{2}\stsq\,d\phi^{2}\,, \label{eq:tdomschw}
\end{equation}
with the coordinate ranges $0 < \tau < 2M$, $\chi_{1} < \chi < \chi_{2}$, $0 < \theta < \pi$, $0 \leq \phi < 2\pi$. Although line element (\ref{eq:tdomline}) can be obtained from (\ref{eq:rdomline}) rather trivially, an interesting complication arises in $f(T)$ gravity which does not occur in the corresponding curvature theories.

Due to the switching of the nature of space and time, a rotation in the interior of the black hole does not generally directly correspond to a rotation in the exterior region. Similarly a boost in the interior does not correspond to a boost in the exterior. One cannot simply take the tetrad which works in the exterior, tetrad (\ref{eq:rdomrottet}), and utilize it in the interior via a simple change of coordinate roles $t \rightarrow \chi$, $r\rightarrow \tau$ along with the switching of the zeroth and first components of the tetrad matrix. Mathematically this is due to the fact that there is no direct analytical extension of coordinate chart (\ref{eq:rdomline}) to coordinate chart (\ref{eq:tdomline}). They are distinct charts despite their similarity.

At this stage there are two choices which will allow us to study the black hole interiors. One choice is to switch to a chart which penetrates the horizon. The advantage will be that one will then have a tetrad which can describe both the exterior and the interior of a black hole. The other choice is to attempt to construct from scratch an acceptable tetrad which can describe the black hole interior. Regarding the first choice, a tetrad we can construct is one compatible with a Painlev\'{e}-Gullstrand type coordinate chart. In the case of $r$ only dependence an acceptable transformation is given by 
\begin{equation}
\tpg=t+\int\stuff\,dr\,,\label{eq:pgtransfa}
\end{equation}
with $P_{0}$ a constant. The transformation matrix can readily be formed and applied to (\ref{eq:rdomrottet}) yielding
\begin{equation}
\left[h^{a}_{\;\mu}\right]_{\mbox{\tiny{pg}}}= \Scale[0.85]{\left( \begin{array}{cccc}
\Ar & -\stuff\Ar & 0 & 0\\
0 & \Br\st\cph & r\ct\cph & -r\st\sph \\
0 & \Br\st\sph & r\ct\sph & r\st\cph \\
0 & \Br\ct & -r\st & 0
\end{array} \right)}\,. \label{eq:pgrottet}
\end{equation}
This tetrad allows for the study of the interior of the black hole as well as the exterior. It also still yields a symmetric $\mathcal{T}_{\mu\nu}$ which is a requirement for an acceptable tetrad. The major disadvantage is that, since the coordinate system is no longer orthogonal, the equations of motion become rather complicated and also the energy conditions become more complicated to analyze than in an orthogonal system.

The second choice is to find a tetrad which works in the orthogonal coordinates of (\ref{eq:tdomline}) in which the analysis for the energy conditions remains relatively simple. Of course, since they are invariant expressions, the energy condition calculations may be done in any coordinate system. In the following, interiors are considered which are $\tau$ only dependent (``T-spheres'' \cite{ref:ruban1}-\cite{ref:zaslatsph}) as adding $\chi$ dependence presents an extremely complicated scenario in the black hole interior even for relatively simple deviations from $f(T)=T$. This class of black hole is necessarily singular in general relativity, \emph{regardless} of energy conditions (see below in section \ref{sss:nisone}). We begin with the diagonal tetrad for the black hole interior 
\begin{equation}
\left[h^{a}_{\;\mu}\right]_{\mbox{\tiny{diag}}}=\left( \begin{array}{cccc}
\alphat & 0 & 0 & 0\\
0 & \betat & 0 & 0 \\
0 & 0 & \tau & 0 \\
0 & 0 & 0 & \tau\st  \end{array} \right)\,, \label{eq:tdomdiagtet}
\end{equation}
and consider rotations of this tetrad about the local Euler angles in the tangent space via the rotation matrices:
\begin{equation}
 \left[R_{x}\right]=\Scale[0.75]{\left( \begin{array}{cccc}
1 & 0 & 0 & 0\\
0 & 1 & 0 & 0 \\
0 & 0 & \cos\psi & \sin\psi \\
0 & 0 & -\sin\psi & \cos\psi  \end{array} \right)},
\left[R_{y}\right]=\Scale[0.75]{\left( \begin{array}{cccc}
1 & 0 & 0 & 0\\
0 & \cos\vartheta & 0 & \sin\vartheta \\
0 & 0 & 1 & 0 \\
0 & -\sin\vartheta & 0 & \cos\vartheta  \end{array} \right)},
\left[R_{z}\right]=\Scale[0.75]{\left( \begin{array}{cccc}
1 & 0 & 0 & 0\\
0 & \cos\varphi & \sin\varphi & 0 \\
0 & -\sin\varphi & \cos\varphi & 0 \\
0 & 0 & 0 & 1  \end{array} \right)},
 \label{eq:rotmats}
\end{equation}
with $\psi$, $\vartheta$, and $\varphi$ functions of the coordinates. In order to find acceptable forms for these angle functions, we must appeal to the conditions outlined in the introduction which provide the criteria for good vs bad tetrads in $f(T)$ gravity. After some work, our calculations reveal that setting
\begin{equation}
 \psi=0,\, \vartheta=\theta +\frac{\pi}{2}, \, \varphi=\phi \label{eq:intconds}
\end{equation}
will satisfy the criteria that yield acceptable equations of motion (i.e. produces a good tetrad). The resulting tetrad has the form
\begin{equation}
\left[h^{a}_{\;\mu}\right]_{\mbox{\tiny{interior}}}=\left( \begin{array}{cccc}
\alphat & 0 & 0 & 0\\
0 & -\betat\cph\st & \tau\sph & \tau\st\cph\ct \\
0 & \betat\sph\st & \tau\cph & -\tau\st\sph\ct \\
0 & -\betat\ct & 0 & -\tau\sin^{2}\theta \end{array} \right)\,. \label{eq:tdomrottet}
\end{equation}
Having found an acceptable tetrad to describe the interior of the black hole we now proceed to study the possibility of making the black hole interior regular everywhere, ideally with non exotic matter.

\section{{The singularity}}\label{S:singularity}
Even in general relativity the issue of what is a serious singularity is often not straight-forward. For example, by definition the best criteria for the presence of a curvature singularity in curvature theories is that one or more components of the Riemann tensor, when projected into the orthonormal frame, becomes infinite. Although one then has a true curvature singularity, it is not necessarily a physically malignant one. One way to ``measure'' the components of the Riemann tensor is via tidal forces along geodesic paths. However, the equation of geodesic deviation has contractions of the Riemann tensor with vectors tangent to the geodesics as well as a contraction with the geodesic deviation vector. It is possible (albeit generally unlikely) in certain situations that within this contraction the infinities present in the Riemann tensor cancel out, and hence the curvature singularity does not spawn infinite tidal forces and is possibly benign. In a similar vein the Kretschmann scalar, being a complete contraction of the Riemann tensor with itself, may be finite yet certain components of the Riemann tensor could be infinite. One then needs to examine other physically measurable quantities to see if there are singularities manifest there. Common examples of such measurable quantities are the stress-energy tensor components in the orthonormal frame; that is, the energy density and fluxes, the pressures, the shears and other stresses.

Another very common criterion for the presence of a pathological singularity in curvature theories is the inability to extend geodesics for arbitrary values of the affine parameter. This criterion of geodesic extension is crucial to the famous Hawking-Penrose singularity theorems \cite{ref:penhawk}. From these theorems it is known that in general relativity one cannot possess a black hole with everywhere extendable geodesics unless the material sourcing the black hole is exotic in some way.

In the sector of torsion gravity, attempts to classify singularities have been performed in \cite{ref:nashednonsing} for M\"{o}ller's tetrad theory and in \cite{ref:torsionsings2} for a class of Riemann-Cartan theories. Another study is \cite{ref:stringstaticbh} for the case of a Schwarzschild black hole in the teleparallel equivalent of general relativity and in \cite{ref:energycontentbh} the authors study energy conditions in an electromagnetic model for the teleparallel equivalent of general relativity. Other analyses can be found in \cite{ref:torsionsings}, and \cite{ref:nonmindiraccoupling} considers cosmological singularities with Dirac fields. An interesting earlier study is in \cite{ref:torschwmetric} within the Poincar\'{e} gauge theory of gravity. From these studies it seems it is even more difficult to discern what is a serious singularity in theories with torsion than in those that are purely curvature based. Here we attempt to study several singularity criteria in the $f(T)$ theory which seem reasonable to us, and we then proceed to alleviate these singular behaviors. We then examine what properties the matter must possess in order to possibly eliminate the singularities.

In the work here we consider several criteria for the presence of a singularity in $f(T)$ gravity. The most straight-forward quantity to calculate is arguably the torsion tensor, which we project in a local orthonormal frame. Utilizing the tetrad (\ref{eq:tdomrottet}) in (\ref{eq:torsion}) and projecting the components into a local orthonormal coordinate system yields:
\begin{subequations}
\begin{align}
T^{\hat{r}}_{\;\hat{t}\hat{r}}=&-T^{\hat{r}}_{\;\hat{r}\hat{t}}=\frac{\betatdot}{\alphat\betat}\,, \label{eq:orthtor1} \\
T^{\hat{r}}_{\;\hat{\theta}\hat{\phi}}         =&-T^{\hat{r}}_{\;\hat{\phi}\hat{\theta}}=    -T^{\hat{\theta}}_{\;\hat{r}\hat{\phi}}=  T^{\hat{\theta}}_{\;\hat{\phi}\hat{r}}= T^{\hat{\phi}}_{\;\hat{r}\hat{\theta}}=-T^{\hat{\phi}}_{\;\hat{\theta}\hat{\phi}}=\frac{2}{\tau}\,, \label{eq:orthtor2}\\
T^{\hat{\theta}}_{\;\hat{t}\hat{\theta}}=& -T^{\hat{\theta}}_{\;\hat{\theta}\hat{t}}= T^{\hat{\phi}}_{\;\hat{t}\hat{\phi}} =-T^{\hat{\phi}}_{\;\hat{\phi}\hat{t}}=\frac{1}{\tau\alphat}\,, \label{eq:orthtor3}
\end{align}
\end{subequations}
where the overdots indicate derivatives with respect to the $\tau$ coordinate. Immediately it can be seen that the components in (\ref{eq:orthtor2}) cannot be made non-singular regardless of the metric functions\footnote{Similar pathologies exist if one chooses to project the tensor into the gravitational tetrad (\ref{eq:tdomrottet}).}. Therefore, we have a non removable torsion tensor singularity. As stated previously, this does not necessarily mean that the spacetime has a physical singularity. Another quantity we can compute is a torsion invariant\footnote{Spacetime coordinate invariant but not locally Lorentz invariant.} somewhat analogous to the Kretschmann scalar, or various permutations thereof. We calculate one as:
\begin{equation}
 T^{\mu\nu\rho}T_{\mu\nu\rho}=\frac{2}{\tau^{2}\alphatsq\betatsq}\left[\betatdot^{2}\tau^{2}-6\alphatsq\betatsq+2\betatsq\right]\,. \label{eq:torinvariant}
\end{equation}
A series expansion of this quantity about the potential singular point yields: 
\begin{equation}
T^{\mu\nu\rho}T_{\mu\nu\rho}=\frac{4}{\tau^{2}\alphao^{2}}\left(1-3\alphao^{2}\right) - \frac{8\dotalphao}{\tau\alphao^{3}} + \mathcal{O}(\tau^{0})\,, \label{eq:torinvexpand}
\end{equation}
where the subscript $0$ indicates that the quantity is to be evaluated at $\tau=0$.
Note from either (\ref{eq:torinvariant}) or (\ref{eq:torinvexpand}) that it may be possible to make this quantity finite for all values of $\tau$.

Yet another quantity which can be calculated to analyze the situation is the torsion scalar (\ref{eq:T}):
\begin{equation}
T=\frac{2}{\tau^{2}\betat\alphatsq}\left[\alphatsq\betat -\betat-2\tau\betatdot\right]\,, \label{eq:torscal}
\end{equation}
which may be expanded about $\tau=0$ as:
\begin{equation}
 T=\frac{2}{\alphao^{2}\tau^{2}}\left[\alphao^{2}-1\right]-\frac{4}{\alphao^{3}\betao\tau}\left[\alphao\dotbetao-\dotalphao\betao\right] +\mathcal{O}(\tau^{0})\,. \label{eq:torscalexpand}
\end{equation}
From these expressions it can be seen that this quantity too can be made finite for all values of $\tau$. However, before continuing we note that one of the conditions required to make (\ref{eq:torinvexpand}) finite at $\tau=0$, viz. $\alphao^{2}=1/3$, is incompatible with one of the conditions which make (\ref{eq:torscalexpand}) finite at $\tau=0$, $\alphao^{2}=1$.

Moving on to more physical criteria we consider the equations of motion of a free test particle. It is a well known interesting fact that in a Weitzenb\"{o}ck spacetime the force equation on a particle subject only to the gravitational field is given by the geodesic equation of the corresponding spacetime in curvature theory in terms of the Riemann-Christoffel connection \cite{ref:telebook}, $\widetilde{\Gamma}^{\mu}_{\;\alpha\beta}$:
\begin{equation}
 \frac{d^{2}x^{\mu}}{d\lambda^{2}} = -\widetilde{\Gamma}^{\mu}_{\;\alpha\beta} \frac{dx^{\alpha}}{d\lambda}\frac{dx^{\beta}}{d\lambda}\,. \label{eq:geoeq}
\end{equation}
The interpretation in teleparallel gravity is that (\ref{eq:geoeq}) yields the acceleration of the particle subject to the force of gravity. For constant mass particles if this becomes infinite then the gravitational force experienced by the particle is infinite. We parametrize the normalized four-velocity, $u^{\alpha}=\frac{dx^{\alpha}}{d\lambda}$, as
\begin{equation}
 [u^{\alpha}]=\left[ \begin{array}{c}
\cosh(\zeta)\,\alphatinv\\[0.2cm]
 \sinh(\zeta)\cos(\omega)\,\betatinv\\[0.2cm]
\sinh(\zeta)\sin(\omega)\cos(\sigma)\, \tau^{-1}\\[0.2cm]
 \sinh(\zeta)\sin(\omega)\sin(\sigma)\,(\tau\sin\theta)^{-1}\end{array}\right]\,, \label{eq:norm4vel}
\end{equation}
with $\zeta$, $\omega$, $\sigma$ functions of $x^{\mu}(\lambda)$. Considering ``radial'' motion ($\omega=0$) we can calculate the gravitational acceleration (or force per unit mass) as:
\begin{equation}
\left[\frac{d^{2}x^{\hat{\mu}}}{d\lambda^{2}}\right]=\left[ \begin{array}{c}
\frac{1}{\alphatsq\betat}\left[\betatdot\alphat\sinh^{2}(\zeta)+\alphatdot\betat\cosh^{2}(\zeta)\right]\\[0.2cm]
\frac{2\betatdot\sinh(\zeta)\cosh(\zeta)}{\alphat\betat}\\[0.2cm]
0\\[0.2cm]
0\end{array}\right]_{|\mbox{path}}\,, \label{eq:accel}
\end{equation}
from which it can be seen that as long as $\alphat$ and $\betat$ do not vanish for finite value of $\lambda$, and the first derivative of $\alpha$ and $\beta$ remains finite, no infinite force will be experienced by the particle (these are sufficient, though perhaps not necessary conditions). The analysis can be extended to massless particles in a straightforward manner but we do not pursue that here.

Finally, we analyze the physical components of the stress-energy tensor. For the material medium we need to choose the form of the stress-energy tensor to be compatible with the equations of motion that result from (\ref{eq:tdomrottet}). That is, it must be representable by a matrix of type-$I_{b}$ and of Segre characteristic $[1,\,1,\,(1,1)]$. We may express this in a very familiar form as
\begin{equation}
 \mathcal{T}^{\mu}_{\;\nu}=\left[\rho+\pperp\right]u^{\mu}u_{\nu}-\pperp\delta^{\mu}_{\;\nu}+[\ppar-\pperp]w^{\mu}w_{\nu}\,, \label{eq:setens}
\end{equation}
with the restrictions $u^{\mu}u_{\mu}=1$, $w^{\mu}w_{\mu}=-1$ and $u^{\mu}w_{\mu}=0$. The quantity (\ref{eq:setens}) has a three-fold importance. One is that these quantities correspond to the proper energy density ($\rho$) and pressures ($\ppar$, $\pperp$) of the material medium, and are therefore in principle physically measurable quantities.  Secondly, they directly correspond to the equations of motion (\ref{eq:eoms}) via $\rho=\mathcal{T}_{\hat{\tau}\hat{\tau}}=\mathcal{T}^{\tau}_{\;\tau}$, $\ppar=\mathcal{T}_{\hat{\chi}\hat{\chi}}=-\mathcal{T}^{\chi}_{\chi}$ and $\pperp=\mathcal{T}_{\hat{\theta}\hat{\theta}}=\mathcal{T}_{\hat{\phi}\hat{\phi}}=-\mathcal{T}^{\theta}_{\theta}=-\mathcal{T}^{\phi}_{\phi}$. Therefore, by demanding that the gravitational equations of motion do not develop a pathology anywhere in the black hole interior we also satisfy non pathological matter. Finally, the quantities $\rho$, $\ppar$ and $\pperp$ are also coordinate scalar quantities as can be easily checked using (\ref{eq:setens}) and the mentioned restrictions on $u^{\mu}$ and $w^{\mu}$, via: $\mathcal{T}^{\mu\nu}u_{\mu}u_{\nu} = \rho$ and $\mathcal{T}^{\mu\nu}w_{\mu}w_{\nu} = \ppar$. Now, since $\rho$ and $\ppar$ are coordinate scalars, the condition $\mathcal{T}^{\mu}_{\;\,\mu}= \rho -2\pperp - \ppar$ establishes that $\pperp$ is also a scalar quantity.

Explicitly, using (\ref{eq:tdomrottet}) we calculate the equations of motion (\ref{eq:eoms}) as:
\begin{subequations}
\allowdisplaybreaks
\begin{align}
& 4\pi\mathcal{T}^{\tau}_{\;\tau}= 4\pi\rho =  \frac{f(T)}{4} + \frac{\Df}{\beta(\tau)\alpha(\tau)^2\tau^2}\bigg(2\dot{\beta}(\tau)\tau + \beta(\tau)\bigg) \, , \label{eq:edens}\\[0.1cm] 
& 4\pi\mathcal{T}^{\chi}_{\;\chi}=-4\pi \ppar =  \frac{f(T)}{4} - \frac{\Df}{\alpha(\tau)^3\beta(\tau)\tau^2}\bigg(\beta(\tau)\dot{\alpha}(\tau)\tau - \alpha(\tau)\beta(\tau) - \alpha(\tau)\dot{\beta}(\tau)\tau \bigg) \nonumber \\
& \qquad - \frac{4\Ddf}{\alpha(\tau)^5\beta(\tau)^2\tau^4} \bigg(\alpha(\tau)\beta(\tau)\ddot{\beta}(\tau)\tau^2 - \dot{\beta}(\tau)\beta(\tau)\alpha(\tau)\tau - \dot{\beta}(\tau)^2\alpha(\tau)\tau^2 \nonumber \\ 
&\qquad   - \dot{\alpha}(\tau)\beta(\tau)^2\tau - 2\dot{\alpha}(\tau)\dot{\beta}(\tau)\beta(\tau)\tau^2 + \alpha(\tau)^3\beta(\tau)^2 - \alpha(\tau)\beta(\tau)^2\bigg) \, , \label{eq:ppar}\\[0.1cm]
& 4\pi\mathcal{T}^{\theta}_{\;\theta}= 4\pi\mathcal{T}^{\phi}_{\;\phi}=-4\pi \pperp= \frac{f(T)}{4} + \frac{\Df}{2\alpha(\tau)^3\beta(\tau)\tau^2}\bigg(\beta(\tau)\alpha(\tau) - \beta(\tau)\alpha(\tau)^3 - \beta(\tau)\dot{\alpha}(\tau)\tau \nonumber \\
& \qquad - \dot{\alpha}(\tau)\dot{\beta}(\tau)\tau^2 + \alpha(\tau)\ddot{\beta}(\tau)\tau^2 + 3\alpha(\tau)\dot{\beta}(\tau)\tau\bigg) + \frac{2\Ddf}{\alpha(\tau)^5\beta(\tau)^3\tau^4} \bigg(\alpha(\tau)\beta(\tau)^3 - \alpha(\tau)^3\beta(\tau)^3 \nonumber \\
& \qquad + 2\beta(\tau)\dot{\alpha}(\tau)\dot{\beta}(\tau)^{2}\tau^{3} - \alpha(\tau)^3\beta(\tau)^2\dot{\beta}(\tau)\tau + \alpha(\tau)\dot{\beta}(\tau)^3\tau^3 - \alpha(\tau)\beta(\tau)\dot{\beta}(\tau)\ddot{\beta}(\tau)\tau^3\nonumber \\[0.1cm]
& \qquad  + 2\alpha(\tau)\beta(\tau)^2\dot{\beta}(\tau)\tau + 2\alpha(\tau)\beta(\tau)\dot{\beta}(\tau)^2 \tau^{2}- \alpha(\tau)\beta(\tau)^2\ddot{\beta}(\tau)\tau^2 + \beta(\tau)^3\dot{\alpha}(\tau)\tau \nonumber \\[0.1cm]
& \qquad   + 3\beta(\tau)^2\dot{\alpha}(\tau)\dot{\beta}(\tau)\tau^{2}\bigg) \,. \label{eq:pperp}
\end{align}
\end{subequations}
These gravitational equations will be used to study the interior region of a time dependent black hole.

\subsection{Energy conditions}\label{S:econd}
Originally the energy conditions were seen as reasonable conditions that all physical (and classical) matter was expected to obey and were independent of theories of gravity \cite{ref:syngebook}. These conditions impose restrictions on the matter such as positivity of the energy density as measured by all causal observers, no superluminal energy transport, etc. In the scenarios given by a diagonal stress-energy tensor and of the type (\ref{eq:setens}) these conditions can be summarized as:
\begin{subequations}
\begin{align}
& \mbox{weak energy condition (WEC):} \qquad \rho \geq 0,\; \rho + \ppar \geq 0,\; \rho + \pperp \geq 0\,, \label{eq:wec} \\[0.2cm] 
& \mbox{dominant energy condition (DEC):} \qquad  \rho - |\ppar| \geq 0,\; \rho - |\pperp|\geq 0\,, \label{eq:dec}\\[0.2cm] 
& \mbox{strong energy condition (SEC):} \qquad  \rho + \ppar \geq 0,\; \rho + \pperp \geq 0,\; \rho+\ppar+2\pperp \geq 0 \,. \label{eq:sec}
\end{align}
\end{subequations}
These conditions will be employed in the analysis below.

\subsubsection{$f(T)=T$}\label{sss:nisone}
This case corresponds to the teleparallel equivalent of general relativity and therefore it is not expected that this scenario will respect energy conditions and at the same time yield a non-singular black hole. We briefly discuss this case here to show how it fails and as a segue to more complicated Lagrangians. Calculating the stress-energy from the equations of motion (\ref{eq:eoms}) yields
\begin{subequations}
\begin{align}
4\pi\rho=&\frac{1}{\betat\alphatsq\tau^{2}}\left[2\betatdot\tau +\betat +\alphatsq\betat\right]\,, \label{eq:n1rho} \\
4\pi\ppar=&-\frac{1}{2\alpha^{3}(\tau)\tau^{2}} \left[\alphat-2\alphatdot\tau +\alpha^{3}(\tau)\right]\,, \label{eq:n1ppar}  \\
4\pi\pperp=&\frac{1}{2\betat\alphatcubed}\left[\alphatdot\betat+\betatdot\alphatdot\tau -\alphat\betatdot -\alphat\betatdotdot\tau\right]\,. \label{eq:n1pperp}
\end{align}
\end{subequations}

Forming the energy conditions (\ref{eq:wec})-(\ref{eq:sec}) and expanding about $\tau=0$ gives
\begin{subequations}
\allowdisplaybreaks
\begin{align}
&4\pi\rho =\frac{1}{2\alphao^{2}\tau^{2}}\left[1+\alphao^{2}\right]+\frac{1}{\betao\alphao^{3}\tau}\left[\alphao\dotbetao-\dotalphao\betao\right] + \mathcal{O}(\tau^{0})\,, \label{eq:n1wec1} \\
&4\pi[\rho+\ppar]= \frac{1}{\alphao^{3}\betao\tau}\left[\alphao\dotbetao+\dotalphao\betao\right] + \mathcal{O}(\tau^{0})  \label{eq:n1wec2} \\
&4\pi[\rho+\pperp]=  \frac{1}{2\alphao^{2}\tau^{2}}\left[1+\alphao^{2}\right] +\frac{1}{2\alphao^{3}\betao\tau} \left[\alphao\dotbetao-\dotalphao\betao\right] + \mathcal{O}(\tau^{0})\,, \label{eq:n1wec3} \\
&4\pi[\rho-|\ppar|\,]=  \frac{1}{\betao\alphao^{3}\left(\alphao^{2}+1\right)\tau}\left[\dotalphao\betao\alphao^{2}+\dotalphao\betao+\alphao^{3}\dotbetao+\alphao\dotbetao    \right] \nonumber \\
&\qquad \qquad \qquad +\mathcal{O}(\tau^{0})\,, \label{eq:n1dec1} \\ 
& 4\pi[\rho-|\pperp|\,]= - \frac{1}{2\alphao^{2}\tau^{2}}\left[1+\frac{1}{\alphao^{2}}\right] +\mathcal{O}(\tau^{-1}) \,, \label{eq:n1dec2} \\
&4\pi[\rho+\ppar+2\pperp] =  \frac{2\dotalphao}{\alphao^{3}\tau} +\mathcal{O}(\tau^{0})\,. \label{eq:n1sec}
\end{align}
\end{subequations}
where the subscript $(0)$ indicates that the quantity is to be calculated at $\tau=0$. Note that sufficiently close to $\tau=0$ it is \emph{not} possible to eliminate the singular terms in these energy conditions\footnote{Inner horizons are not considered for T-spheres.}. Since the energy conditions are comprised of linear combinations of $\rho$, $\ppar$, and $\pperp$, this implies it is not possible to eliminate singularities in at least some of these physical quantities. In order to do so one needs to abandon an everywhere Lorentzian spacetime, and a Riemannian region (-4 signature) is required instead of a pseudo-Riemannian spacetime in the vicinity of $\tau=0$. Therefore the gravitational equations of motion, being singular, are not valid at $\tau=0$. These results are consistent with general relativity as expected for $f(T)=T$. (Specifically, they are problematic from a singularity elimination point of view as can be seen in the lowest order terms in (\ref{eq:n1wec1}), (\ref{eq:n1wec3}) and (\ref{eq:n1dec2}).)

\subsubsection{Comments for general $n$}
Substituting the desired function for $f(T)$, $f(T) = T+b_nT^n$, and series expanding the components of $\mathcal{T}_{\mu}^{\,\,\,\nu}$ around $\tau = 0$, we find that for negative values of $n$ the qualitative singularity situation does not change in comparison with the general relativity scenario ($n=1$). That is, for negative $n$ various components of the mixed stress-energy tensor have singular terms at $\tau=0$ with expressions such as $1+\alphao^{2}$ in the numerator. This prohibits singularity resolution again unless a Riemannian region is allowed. The reason there is no change for negative powers of $n$ in comparison to general relativity is because of the way $f(T)$ enters into the equations of motion (\ref{eq:eoms}); negative powers of $n$ yield higher-order (in $\tau$) corrections to $\rho$, $\ppar$, and $\pperp$. That is, the $n=1$ term is the term which contributes to lowest power in $\tau$ when considering supplements with negative $n$.

For positive $n$ the leading order singular terms are of order $\mathcal{O}(\tau^{-2n})$. We further note that in our calculations we find that the conditions $\alphao = 1$ and $\betao = \dotbetao/\dotalphao$ are always necessary to negate the leading order singular terms. These are the same conditions required to regularize the torsion scalar (\ref{eq:torscal}), so non-singular stress-energy components necessitate a non-singular torsion scalar. This, of course, is related to the fact that the torsion scalar appears directly in the equations of motion.

At this point, finding additional conditions with the components of the stress-energy tensor for general $n$ becomes prohibitively difficult. However the goal is to show that for $n \neq 1$ it is possible to have regular black holes, so we concentrate on several powers of $n$ to resolve the issue. Specifically we report below on $n = 2$, although we have also studied all the cases $-15 \leq n \leq 7$, and all positive $n$ can be made to eliminate singularities in the physical stress-energy and the torsion scalar (\ref{eq:torscal}), and hence the gravitational equations of motion are analytic throughout the manifold for $n>1$, as well as render forces finite in the force equation (\ref{eq:accel}). The WEC and SEC can be respected in all the positive $n$ scenarios.

\subsubsection{{$f(T)=T+b_{2}T^{2}$}}

For the $n=2$ case we substitute $f(T) = {T} + b_2{T}^2$ in (\ref{eq:gravact}) and calculate the subsequent equations of motion (\ref{eq:eoms}). Series expanding the field equations around $\tau = 0$ allows us to determine the structure of the matter in the vicinity of the possible singular point. As noted previously, the $\mathcal{O}(\tau^{-4})$ and $\mathcal{O}(\tau^{-3})$ terms in all the expanded components of the stress-energy tensor are negated with the conditions $\alphao = 1$ and $\betao = \dotbetao/\dotalphao$. In order to negate the $\mathcal{O}(\tau^{-2})$ and $\mathcal{O}(\tau^{-1})$ components we must constrain more expansion coefficients. Note that in general the expansion coefficients are free, meaning there are infinitely many possible solutions. In actuality, the metric functions and coupling constants would be constrained experimentally. However, as we are considering a black hole interior an experimental method is not readily apparent. For simplicity, let $\dotalphao = \dotdotalphao = \dotdotdotalphao = \dotdotdotdotalphao = 1$, $\betao = 1$, $\dotdotbetao = 15/8$, and $\dotdotdotbetao = 107/24$, which also enforces $\dotbetao = 1$. We again stress that these are not required conditions, but serve as a specific example out of infinitely many to make the analysis perspicuous. Under these conditions the series expansions are:
\begin{subequations}
\begin{align}
4\pi\mathcal{T}_{\tau}^{\,\,\tau} &= 4\pi\rho = \frac{1 + b_2}{\tau^2} + \frac{1175b_2}{72} - \frac{4b_2\dotdotdotdotbetao}{3} - \frac{1}{8} +\mathcal{O}(\tau)\,, \\
-4\pi\mathcal{T}_\chi^{\,\,\chi} &= 4\pi \ppar = -\left[ \frac{1 + b_2}{\tau^2} - 2\frac{1 + b_2}{\tau} + \frac{1259b_2}{24} - 4b_2\dotdotdotdotbetao + 3 \right] +\mathcal{O}(\tau)\,, \\
-4\pi\mathcal{T}_\theta^{\,\,\theta} &= 4\pi \pperp  = -\frac{1247b_2}{72} + \frac{4b_2\dotdotdotdotbetao}{3} - \frac{7}{8}+\mathcal{O}(\tau)\,.
\end{align}
\end{subequations}
It is immediately clear that if $b_2 = -1$ all of the components are non-singular.

The remaining undefined coefficient present in the above expansions is $\dotdotdotdotbetao$. This final constraint comes from the energy conditions, equations~(\ref{eq:wec}) -~(\ref{eq:sec}). From the weak energy condition, we have:
\begin{center}
\begin{equation}
\dotdotdotdotbetao - \frac{37}{3} \geq 0\,\,\text{ and }\,\, \frac{4754}{384} - \dotdotdotdotbetao \geq 0.
\label{eq:beta4_WEC}
\end{equation}
\end{center}
Note that satisfying equation~(\ref{eq:beta4_WEC}) also satisfies the null energy condition. The strong energy condition puts a further constraint on the range of $\dotdotdotdotbetao$:
\begin{center}
\begin{equation}
\frac{4745}{384} - \dotdotdotdotbetao \geq 0.
\label{eq:beta4_SEC}
\end{equation}
\end{center}
So the range of values of $\dotdotdotdotbetao$ that will satisfy the weak, null, and strong energy conditions is
\begin{center}
\begin{equation*}
\frac{37}{3} \leq \dotdotdotdotbetao \leq \frac{4745}{384} < \frac{4754}{384}.
\end{equation*}
\end{center}
However the dominant energy condition at $\tau=0$ selects one value out of this range. The dominant energy condition is satisfied only if $\dotdotdotdotbetao = \frac{4745}{384}$, which enforces $4\pi\rho = 4\pi\pperp = 1/32$ and $4\pi \ppar = -1/32$. So the metric functions that satisfy the energy conditions at $\tau=0$ in this particular case are:
\begin{subequations}
\begin{align}
\alpha(\tau) &= 1 + \tau + \frac{1}{2}\tau^2 + \frac{1}{6}\tau^3 + \frac{1}{24}\tau^4 +\mathcal{O}(\tau^{5})\,, \nonumber \\
\beta(\tau) &= 1 + \tau + \frac{15}{16}\tau^2 + \frac{107}{144}\tau^3 + \frac{4745}{9216}\tau^4 +\mathcal{O}(\tau^{5})\, . \nonumber
\end{align}
\end{subequations}
We note that the above solution has non-zero neighborhoods about $\tau=0$ which respect the WEC and SEC, but although the DEC is satisfied at $\tau=0$ it is violated in a neighborhood as one moves away from $\tau=0$. Attempts at a general analysis proved difficult, but no values of parameters which were studied yielded a non-zero DEC respecting neighborhood about $\tau=0$, although it was found that the region of DEC violation could be made very small.

The above can be treated as a local specific solution (one of many allowed which work). Away from $\tau=0$ one can patch the solution to energy condition respecting, non-singular solutions. It is not difficult to do so but there is the issue of appropriate junction conditions in $f(T)$ gravity. In our tests we employed Synge's junction condition \cite{ref:syngebook} as it is a condition derived on the matter field and does not require detailed knowledge of the particular gravitational equations. This condition, $\left[\mathcal{T}^{\mu}_{\;\nu} \widehat{n}_{\mu}\right]_{|\pm \tau=J(\chi)}$ ($\widehat{n}_{\mu}$ being a unit normal covector to the junction surface $\tau=J(\chi)$) for the spherically symmetric T-domain may be summarized as 
\begin{subequations}
\begin{align}
\left[\mathcal{T}^{\chi}_{\;\chi}\,\partial_{\chi}J(\chi) - \mathcal{T}^{\tau}_{\;\chi}\right]_{|\pm\;\tau=J(\chi)}=0\,, \label{eq:sycond1} \\
\left[\mathcal{T}^{\chi}_{\;\tau}\, \partial_{\chi}J(\chi) - \mathcal{T}^{\tau}_{\;\tau}\right]_{|\pm\;\tau=J(\chi)}=0\,, \label{eq:sycond2}
\end{align}
\end{subequations}
where the subscript $\pm$ indicates that we are considering the discontinuity in the quantity in square brackets on the junction surface $\tau=J(\chi)$. Given that our scenario is $\chi$ independent, yielding a diagonal stress-energy tensor, these conditions boil down to continuity of $\mathcal{T}^{\tau}_{\;\tau}$ at the junction surface $\tau=J_{0}=\mbox{const.}$, which we find can be easily satisfied. In fact, we find that an even stronger condition can be met where derivatives of the tetrad functions up to arbitrary order may be made continuous across the junction. We therefore find that it is possible to respect energy conditions in regular black holes of this type within $f(T)$ theories, save for a small region near $\tau=0$. 

For cases with $2 \leq n \leq 7$ it is also possible to construct metric functions that regularize the black hole and describe matter that satisfy the energy conditions as above. Note that for $n$ outside this range analysis becomes prohibitively difficult, due to the length of the expressions in the various expansions. We can say that it is necessary in general that $b_n < 0$. The expansion coefficients of $\alpha(\tau)$ and $\beta(\tau)$ are generally unrestricted, with the exception of $\alphao$ and $\betao$. However, artificially restricting the derivatives of $\alphat$ will always generate a working (non-singular) solution in the same manner as shown for $n = 2$. The conclusion is that for any $f(T) = T + b_nT^n$, $2 \leq n \leq 7$, singularities which are necessarily present in general relativity for this class of black hole, can be alleviated by matter that respects the WEC and SEC, although there is still a price to pay in that the DEC is violated in a small region about $\tau=0$ (though it need not be violated right at $\tau=0$). We conjecture that this is true for all positive integer values of $n$, and hence there may exist a singularity theorem for the positive $n$ sector for T-sphere black holes in the case $n>1$. As discussed above, the same is not true for the negative sector of $n$. For the negative sector the spacetime remains necessarily singular as in general relativity or its teleparallel equivalent $f(T)=T$.

\section{Concluding Remarks}\label{S:conclusion}
A tetrad has been derived which is suitable for describing the interior of a class of spherically symmetric black holes, which are necessarily singular in general relativity, in the extended teleparallel theory of gravity. This tetrad allows for the study of potentially singular quantities inside the black hole. Specifically, several criteria for singularities were considered and it was found that although the orthonormal torsion tensor cannot be made finite everywhere, finite torsion gravitational forces can be ensured by demanding finite non-zero metric functions near the potentially singular points. As well, for Lagrangians which consist of various powers of the torsion scalar, it is shown that the matter field quantities remain non-singular, unlike the case in general relativity, and hence the gravitational equations of motion remain valid throughout the black hole interior manifold. (Extensions to negative values of $\tau$ are in principle allowed.) At the same time, all energy conditions considered can be satisfied inside the black hole where the general relativity singularity occurs, although we find that the dominant energy condition is violated in an arbitrarily small neighborhood about this point. This leads us to speculate that a singularity theorem holds for these black holes in the extended teleparallel gravity. 

The above conditions on the matter also ensure that the torsion scalar is finite. Although scalars created out of the torsion are scalars under general coordinate transformations, in the torsion theory statements regarding these quantities are not locally Lorentz invariant. They are however globally Lorentz invariant, and for local Lorentz transformations these quantities, although they will change, will not become singular as long as the local Lorentz transformation is not singular. Hence, regular black holes of this type are permitted while preserving the weak energy and strong energy conditions everywhere, and the dominant energy condition almost everywhere. It was found these results hold for all extensions to the teleparallel equivalent of general relativity studied as long as the powers of $n$ are positive. This is not true for negative powers. It seems likely therefore that torsion gravitational Lagrangians which are Taylor expandable allow for regular black holes in cases where general relativity does not and that the matter can obey the WEC and SEC with a minor violation of the DEC. Singularities are therefore easier to remedy within $f(T)$ theory while still retaining second-order field equations, which is not afforded by most curvature extensions of gravity. 

The analysis presented here can be easily extended to non-Laurent or Taylor expandable Lagrangians. There are a number of interesting studies in the $f(T)$ literature regarding the ability of extended teleparallel gravity to successfully produce the observed acceleration of the universe \cite{ref:cosmocardone}-\cite{ref:cosmobamba}, including recent extensions to anisotropic models \cite{ref:ksrahaman}. It would be interesting to see if the same Lagrangians which are capable of yielding the observed cosmological acceleration are also capable of eliminating the big-bang and black hole singularities without the need to resort to exotic matter.

\vspace{-0.15cm}
\section*{Acknowledgments}
We would like to thank D. Horvat and S. Iliji\'{c} of the University of Zagreb and K.S. Viswanathan of Simon Fraser University for a number of very helpful discussions.




\vspace{0.5cm}
\renewcommand{\refname}{{References}}
\linespread{0.6}
\bibliographystyle{unsrt}

\end{document}